\documentclass[useAMS,usenatbib,usegraphicx]{mn2e}

\def\simlt{\lower.5ex\hbox{$\; \buildrel < \over \sim \;$}}
\def\simgt{\lower.5ex\hbox{$\; \buildrel > \over \sim \;$}}

\def\Lya{\mbox{Ly$\alpha$} }
\def\Lyb{\mbox{Ly$\beta$} }
\def\Lyc{\mbox{Ly$\gamma$} }
\def\bk{{\bf k}}
\def\tot{\mathrm{tot}}

\def\rt{\tilde{r}}

\def\td{{\tilde{\delta}}}
\def\beq{\begin{equation}}
\def\eeq{\end{equation}}
\title{Detecting Early Galaxies Through Their 21-cm Signature}

\author[Smadar Naoz, and Rennan Barkana]{Smadar Naoz$^{1}$ and 
Rennan Barkana$^{1}$\thanks{E-mail: smadar@wise.tau.ac.il (SN); barkana@wise.tau.ac.il 
(RB)}\\ $^{1}$School of Physics and Astronomy, The Raymond and Beverly
Sackler Faculty of Exact Sciences,\\ Tel Aviv University, Tel Aviv
69978, ISRAEL}

\begin{document}

\pagerange{\pageref{firstpage}--\pageref{lastpage}} \pubyear{2007}
\maketitle
\label{firstpage}

\begin{abstract}
  New observations over the next few years of the emission of distant
  objects will help unfold the chapter in cosmic history around the
  era of the first galaxies. These observations will use the neutral
  hydrogen emission or absorption at a wavelength of 21-cm as a
  detector of the hydrogen abundance. We predict the signature on the
  21-cm signal of the early generations of galaxies. We calculate the
  21-cm power spectrum including two physical effects that were
  neglected in previous calculations. The first is the redistribution
  of the UV photons from the first galaxies due to their scattering
  off of the neutral hydrogen, which results in an enhancement of the
  21-cm signal. The second is the presence of an ionized hydrogen
  bubble near each source, which produces a cutoff at observable
  scales. We show that the resulting clear signature in the 21-cm
  power spectrum can be used to detect and study the population of
  galaxies that formed just 200 million years after the Big Bang.
\end{abstract}


\section{Introduction}\label{intro}

An important milestone in the evolution of the Universe is the
appearance of the first luminous objects, which begins the era of
heating and ionization of the intergalactic medium (IGM), referred to
as cosmic reionization. A promising probe of the cosmic gas up to the
end of reionization is the hyperfine spin-flip transition of neutral
hydrogen (H~I) at a wavelength of 21-cm. Observations of the
redshifted 21-cm power spectrum as a function of wavelength and
angular position can provide a three-dimensional map of the
H~I distribution \citep{H_R,madau}.

The 21-cm line of the hydrogen atom results from the transition
between the triplet and singlet hyperfine levels of the ground state.
Calculations of 21-cm absorption begin with the spin temperature
$T_s$, defined through the ratio between the number densities of
hydrogen atoms, ${n_1}/{ n_0}=3\exp(-T_\star/T_s)$, where subscripts
$1$ and $0$ correspond to the excited and ground state levels of the
21-cm transition and $T_\star=0.0682$~K corresponds to the energy
difference between the levels. The 21-cm spin temperature is on the
one hand radiatively coupled to the cosmic microwave background (CMB)
temperature $T_\gamma$, and on the other hand coupled to the kinetic
gas temperature $T_k$ through collisions \citep{AD} or the absorption
of \Lya photons, as discussed below. For the concordance set of
cosmological parameters \citep{Spergel06}, the mean brightness
temperature on the sky at redshift $z$ (relative to the CMB itself) is
$\bar{T}_b = 28\, {\rm mK}\,\sqrt{(1+z)/10}\, [(T_s-
T_{\gamma})/T_s]\, \bar{x}_{\rm HI}\,$ \citep{madau} where
$\bar{x}_{\rm HI}$ is the mean neutral fraction of hydrogen.

After cosmic recombination, down to $z\sim 200$ the remaining free
electrons kept $T_k$ close to $T_\gamma$. Afterward, the gas cooled
adiabatically and atomic collisions kept the spin temperature $T_s$
coupled to $T_k$, both of them lower than $T_\gamma$, thus creating a
21-cm signal in absorption. As the universe continued to expand and
rarefy the gas, the radiative coupling between $T_\gamma$ and $T_s$
started to dominate again, and the 21-cm signal faded. Starting at
$z\sim 66$ \citep{NNB}, the UV photons from the first galaxies emitted
between the \Lya and Lyman limit wavelengths propagated freely in the
IGM. They redshifted or scattered into the \Lya resonance, thus
coupling $T_s$ to $T_k$ through the WF effect \citep{Wout,field}
whereby atoms re-emitting \Lya photons can de-excite into either of
the hyperfine states.

Fluctuations in the \Lya radiation emitted by the first galaxies led
to fluctuations in the 21-cm signal \citep{BL05b}, until the WF effect
saturated (i.e., finished dropping $T_s$ to $T_k$). In particular, the
biased fluctuations in the number density of galaxies, combined with
Poisson fluctuations in the number of galaxies and the $1/r^2$
dependence of the flux, caused large fluctuations in the \Lya
background. This \Lya background is composed of two parts, the photons
that directly redshift into the \Lya resonance, and those produced
during the atomic cascade from higher Lyman series photons (\Lyc and
above). About $30\%$ of the total number of photons emitted from the
first stars between \Lyc and the Lyman limit are converted to \Lya
\citep{Jonathan06a,Hirata}. The 21-cm power spectrum that arises from
the fluctuations in the \Lya background can be used to probe the
first sources of light and their effect on the IGM
\citep{madau,Miralda,BL05a,BL05b,Shapiro}.

In this \emph{Letter}\/ we focus on two effects that have not
previously been included in calculations of statistics of the 21-cm
brightness temperature; these effects dramatically influence the \Lya
flux fluctuations and therefore the 21-cm power spectrum. The first is
the diffusion of photons through the neutral hydrogen gas. Since the
optical depth of \Lya is around a million, photons scattered on the
surrounding gas cause each source to appear as a halo in the sky
rather than a point source \citep{LR99}. In the absence of scattering
a photon will redshift to \Lya at some distance from the source, but
in the presence of scattering the emitted photons scatter back and
forth and thus reach \Lya much closer to the source
\citep{Leonid07b,Semelin}. Thus, the \Lya flux that a random point in
the universe observes is dominated by nearby sources. As a result, the
fluctuations in the \Lya flux are larger and enhance the 21-cm power
spectrum. The second effect is caused by the presence of ionized
hydrogen (H~II) around each source \citep[e.g.,][]{BL01}. In this
region photons can redshift below the \Lya resonance without
encountering H~I, and thus not participate in the WF effect. Thus, a
typical H atom (which by definition is not within an H~II region)
receives no \Lya photons from sources closer than the typical size of
an H~II bubble. Since galaxy number fluctuations below this scale do
not contribute to \Lya fluctuations, we expect a cutoff in the 21-cm
power spectrum around this scale.

Our calculations are made in a $\Lambda$CDM universe matching
observations \citep{Spergel06}, with a power spectrum normalization
$\sigma_8=0.826$, Hubble constant $H_0=68.7 \mbox{ km
s}^{-1}\mbox{ Mpc}^{-1}$, and density parameters $\Omega_m=0.299$,
$\Omega_\Lambda=0.701$, and $\Omega_b=0.0478$ for matter, cosmological
constant, and baryons, respectively.

\section{Lyman series scattering}

We assume a uniform neutral IGM that exhibits pure Hubble expansion
around a steady point source \citep{LR99}. The surrounding IGM, mainly
H~I, is optically thick to the UV radiation that the source emits. We
use a Monte-Carlo method for the scattering \citep{LR99}, for photons
of various initial frequencies between \Lya and the Lyman limit. The
scattering redistributes the distances at which the photons transform
to Ly$\alpha$, affecting strongly the photons originally emitted
between \Lya and \Lyb. Due to flux conservation, the enhancement of
the flux at small scales (by up to a factor of $\sim 5$ compared to
the optically thin case) is balanced by a steep drop on large scales
(where $\nu\to$Ly$\beta$). \Lya photons produced via cascading from
photons emitted above \Lyc are much less affected (see below), but we
fully include these cascades assuming that the resulting \Lya photons
are emitted at the line center \citep{FJ06}. We account for several
features that substantially affect the redistribution of photons
compared to previous calculations that gave a divergent $r^{-1/3}$
enhancement near the source
\citep{Leonid07b,Semelin}. First, we explicitly include the H~II
region around each source, which results in enhanced \Lya scattering
just outside the H~II region; the enhancement has a shallower rise to
a peak value and then drops to zero right near the H~II region,
because of the loss of photons that are scattered back into the H~II
region. Second, each hydrogen atom receives \Lya photons from sources
with an effective spectrum that drops sharply at frequencies
approaching Ly$\beta$; photons at such frequencies are emitted by
time-retarded sources that formed in a universe with fewer galaxies,
effectively reducing the emission rate (see also section~4 below). We
find that scattering strongly enhances this drop in the
spectrum. Finally, we also account for the variation of the Hubble
constant with redshift, which substantially affects the photons
emitted at $\nu\to$Ly$\beta$.

\begin{figure}%[th]
\includegraphics[width=84mm]{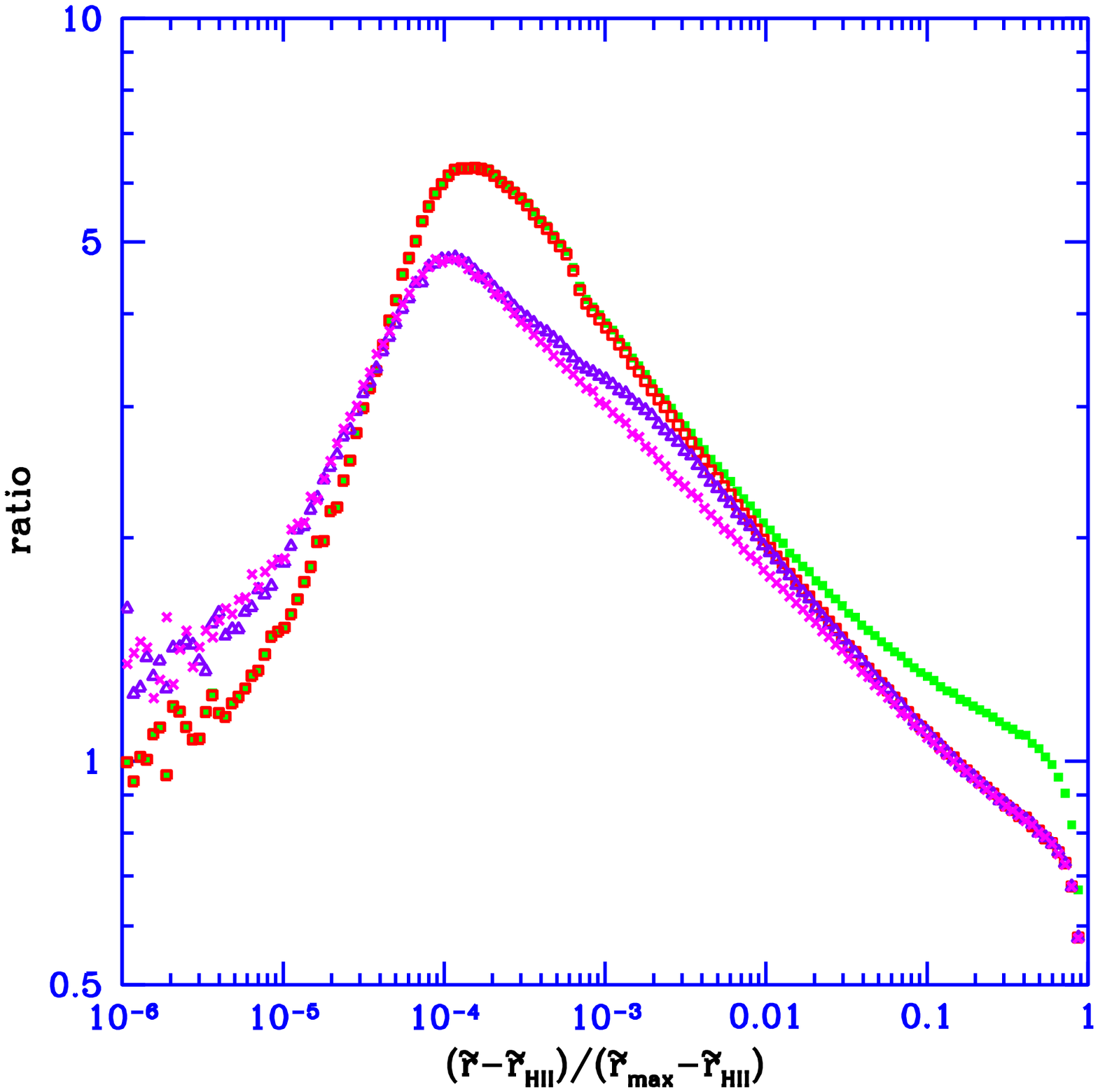}
\caption{The ratio between the flux as a result of our calculation as
  described in the text and the flux for the case with no scattering
  redistribution, for photons emitted between \Lya and Ly$\beta$ at
  $z=20$. We consider an H~II region of size 0.023, 0.23, and 2.3
  comoving Mpc (open squares, triangles, and crosses, respectively).
  Assuming Population II stars, the filled squares (light green) are
  for the case of a simple power-law spectrum (with $r_{\rm
  HII}=0.023$ Mpc), while the other curves assume the correct
  effective spectrum with time-retarded sources. For the effective
  spectrum, the flux is low at large distances, so the scattering
  (which conserves the total photon number) must lower this flux
  especially strongly in order to compensate for the enhancement at
  small distances.}
\label{ratio}
\end{figure}

Following \citet{LR99} we define a dimensionless variable
$\rt={r}/{r_\star}$, where $r_\star$ is the radius for which
redshifting a \Lya photon due to Hubble expansion would produce a
frequency with a remaining optical depth of unity (from there out to
infinity, in the direction away from the line center). For example,
using the standard cosmological parameters, $r_\star\sim$ 1 Mpc
(physical), independent of redshift (for a source at high redshift:
\citet{LR99}). In Figure~\ref{ratio} we show the ratio between the
flux of the source as a result of our calculation as described above
and the flux for the case with no redistribution due to scattering,
i.e., where the photons simply redshift until they are absorbed at
line center. For photons emitted between \Lya and \Lyb we have plotted
the ratio as a function of the variable $\tilde{x}=(\rt-\rt_{\rm
HII})/(\rt_{\rm max}-\rt_{\rm HII})$, where $r_{\rm max}$ is the
radius at which an emitted \Lyb photon redshifts all the way to
Ly$\alpha$, and the various radii are expressed in units of $r_\star$.
At $z=20$, $r_\star=22.8$ comoving Mpc and $r_{\rm max}=321.4$
comoving Mpc.  Showing the results as a function of the variable
$\tilde{x}$ provides a better view of the behavior of the flux very
near to the H~II region\footnote{Note that the numerical noise seen in
the Figure very near the H~II region occurs where the flux is not
significantly enhanced, and thus does not affect our results.
Furthermore, this region is within the thickness of the shell of the
H~II region, and thus is even less important in practice.}. We have
considered in this Figure a Population II spectrum source, i.e., we
approximated the emissivity within this frequency region as $\approx
\nu^{\alpha_s}$, where for Population II $\alpha_s=0.14$
\citep{BL05b,Bromm01}. We have also performed the calculation for a
spectrum appropriate for Population III stars (see below). For photons
cascading from higher-order Lyman lines, the atomic constants would
make the enhancement occur much closer to the H~II region edge; e.g.,
even for Ly$\gamma$, $r_\star=0.152$ comoving Mpc, $r_{\rm max}=40.3$
comoving Mpc, and a substantial enhancement occurs only at $\tilde{x}
\simlt 10^{-5}$ (with the corresponding definition of $\tilde{x}$ for
Ly$\gamma$), well within the thickness of the partial ionization shell
of the H~II region, where the flux is in practice strongly suppressed.

\section{21cm Fluctuations}% and the Spin Temperature}

In general, fluctuations in $T_b$ can be sourced by fluctuations in
the gas density, temperature, neutral fraction, radial velocity
gradient, and \Lya flux from galaxies. To linear order, perturbing the
equation of the brightness temperature, we have $\delta T_b=\beta_b
\delta_b+\beta_T\delta_T+\delta_{x}+\beta_\alpha\delta_\alpha-
\delta_{d_rv_r}$, where each $\delta_i$ describes the fractional
perturbation on the quantity $i$; $b$ denotes the baryon density, $T$
the gas temperature, $x$ the neutral fraction, $\alpha$ the \Lya
coefficient $x_\alpha$ (which is simply the \Lya flux measured in
units so that \Lya flux fluctuations give the largest 21-cm
fluctuations when $x_\alpha=1$), and $d_rv_r$ the line of sight
velocity gradient [which creates an apparent density fluctuation
\citep{Indian2,BL05a}]. The associated time-dependent $\beta_i$
coefficients are $\beta_b=1+[x_c/\tilde{x}_{ \tot}]$,
$\beta_\alpha=x_\alpha/\tilde{x}_{ \tot}$,
$\beta_T=[T_\gamma/(T_k-T_\gamma)]+(x_c/\tilde{x}_{ \tot})* [d \log
\kappa_{1-0} / d \log T_k]$, where $\tilde{x}_{ \tot} \equiv
x_{\tot}(1+x_{\tot})$, $x_{\tot} \equiv x_\alpha+x_c$, and $x_c$ and
$\kappa_{1-0}$ describe the collisional 21-cm excitation
\citep{AD,BL05a,F06}. We consider the high-redshift regime before
significant cosmic reionization or stellar heating. Due to Compton
scattering of the CMB photons with the remaining free electrons after
cosmic recombination, the baryon density and temperature fluctuations
are not proportional \citep{NB}.

A Fourier transform yields the brightness temperature fluctuation in
$\bk$-space (where $\tilde{\delta}$ denotes the transform of each
$\delta$ quantity) \citep{BL05a}:
\begin{equation}
\label{Tbk_n}
\tilde{\delta}{T_b} (\bk,t) = 
\left(\mu^2\, r_{[\dot{\delta}:\delta]}+\beta \right)
\tilde{\delta}_b(\bk) + \beta_\alpha
\tilde{\delta}_\alpha(\bk)\ ,
\end{equation}
where the ratio $r_{[\dot{\delta}:\delta]}(k,z) \equiv
[(d/dt)\tilde{\delta}_b]/(H\, \tilde{\delta}_b)$, $\beta(k,z) \equiv
\beta_b+r_{[T:\delta]}\beta_T$ in terms of $r_{[T:\delta]} \equiv
\tilde{\delta}_T / \tilde{\delta}_b$, and $\mu = \cos\theta_k$ in
terms of the angle $\theta_k$ of the wavevector $\bk$ with respect to
the line of sight. %Hereafter we neglect the ionization fluctuations,
%assuming we are considering a time long before reionization, when
%ionization fluctuations were negligible compared to the other sources
%of 21-cm fluctuations.

We denote by $P_{\delta_b}(k)$ and $P_{\alpha}(k)$ the power spectra
of the fluctuations in baryon density and in the \Lya flux,
respectively, and the power spectrum $P_{\delta_b-\alpha}$ as the
Fourier transform of their cross-correlation function \citep{BL05b}.
The 21-cm power spectrum can then be written in the form $P_{T_b}(\bk)
= \mu^4P_{\mu^4}(k)+\mu^2P_{\mu^2}(k)+P_{\mu^0}(k)\,,$ where each of
the three $\mu$ coefficients can be separately measured from their
different $\mu$ dependence \citep{BL05a}. This angular separation of
power makes it possible to detect separately different physical
aspects influencing the 21-cm signal. In terms of our definitions
above, the coefficients are
\begin{eqnarray}
\label{powTb} 
P_{\mu^4}(k) & = & r^2_{[\dot{\delta}:\delta]} P_{\delta_b}(k) \\
P_{\mu^2}(k) & = &  2 r_{[\dot{\delta}:\delta]} \left[ \beta 
P_{\delta_b}(k) + \beta_\alpha P_{\delta_b-\alpha}(k) \right] 
\nonumber \\  P_{\mu^0}(k) & = & \beta^2 P_{\delta_b}(k) + 
\beta^2_\alpha P_{\alpha}(k) + 2 \beta \beta_\alpha 
P_{\delta_b-\alpha}(k)\ . \nonumber 
\end{eqnarray}
It is possible to probe whether some sources of $P_{\alpha}$ are
uncorrelated with $\delta_b$; sources that are linear functionals of the baryon
density distribution do not contribute to 
the following quantity \citep{BL05a}:
\beq P_{\rm un-\delta}(k) \equiv P_{\mu^0}- \frac{1}{4} \frac{
P_{\mu^2}^2} {P_{\mu^4}} = \beta^2_\alpha \left( P_\alpha -
\frac{P_{\delta_b-\alpha}^2} {P_{\delta_b}} \right)\ . \eeq 

\section{The \Lya Flux of Galaxies}\label{sec:flux}

We summarize here the calculation of the \Lya flux \citep{BL05b}. The
intensity of \Lya photons observed at a given redshift is due to
photons that were originally emitted between the rest-frame
wavelengths of \Lya and the Lyman limit. Photons that were emitted
below \Lyb by a source simply redshift until they are seen by an atom
at $z$ at the \Lya wavelength. Such photons can only be seen out to a
distance corresponding to the redshift $z_{\rm max}(2)$, where
$[1+z_{\rm max}(2)]/(1+z) = (32/27)$, the ratio of \Lya to \Lyb
wavelengths. Photons above the \Lyb energy redshift until they reach
the nearest atomic level $n$. The neutral IGM is opaque even to the
higher levels and so the photon is repeatedly scattered, with a $\sim
30\%$ chance of being downgraded to a \Lya photon and continuing to
scatter, except that the chance is zero for a \Lyb photon
\citep{Jonathan06a,Hirata}. To be seen at the \Lya resonance at $z$,
the photon must have been emitted below a maximal redshift $z_{\rm
max}(n)$. The intensity is then \citep{BL05b}:
\begin{eqnarray}
 \label{eq:Ja} 
J_{\alpha}&=&\frac{(1+z)^2} {4 \pi}\times\\
&& \sum_{n=2}^{n_{\rm max}} f_{\rm
recyc}(n) \int_{z_{\rm H\,II}}^{z_{\rm max}(n)} \frac{c dz'}{H(z')}\,
\epsilon(\nu_n', z')\, f_{\rm scat}(n,r) \nonumber \ ,
\end{eqnarray}
where absorption at level $n$ at redshift $z$ corresponds to an
emitted frequency $\nu_n'$ at $z'$, and $\epsilon$ is the photon
emissivity.
%\beq 
%\nu_n' = \nu_{\rm LL} (1-n^{-2}) {(1+z')\over (1+z)}\ , 
%\label{eq:nu} 
%\eeq 
%in terms of the Lyman limit frequency $\nu_{\rm LL}$. 
We have included the factor $f_{\rm recyc}$ \citep{Jonathan06a,Hirata}
which is the fraction of photons absorbed at level $n$ that are
eventually recycled into \Lya photons in their subsequent cascade. The
new correction factor $f_{\rm scat}$ is the overall factor by which
the \Lya flux is changed due to the scattering-induced redistribution
of photons that are emitted with frequencies between levels $n$ and
$n+1$ from sources at a comoving distance $r$ from the final
destination at redshift $z$ (where $r$ is a function of $z$ and
$z'$). Also new to calculations of the 21-cm power spectrum is the
lower limit of the integral ($z_{\rm H\,II}$), which accounts for the
H~II regions; in order to receive \Lya photons from sources closer
than the typical size of an H~II region, an H atom would have to be
inside an H~II bubble -- a contradiction.

\section{The 21-cm power spectrum}

As mentioned above, there are two separate sources of fluctuations in
the \Lya flux \citep{BL05b}. The first is density inhomogeneities.
Since gravitational instability proceeds faster in overdense regions,
the biased distribution of rare galactic halos fluctuates much more
than the global dark matter density \citep{Kaiser,BL04}. When the
number of sources seen by each atom is relatively small, Poisson
fluctuations provide a second source of fluctuations, statistically
independent of the first source to linear order. Unlike typical
Poisson noise, these fluctuations are correlated between gas elements
at different places, since two nearby elements see many of the same
sources, though at different distances \citep{BL05b}.  Note that
although each hydrogen atom receives some \Lya flux from sources as
far away as 300 comoving Mpc, the fluctuations in flux are actually
relatively large because a significant portion of the flux comes from
nearby sources.

Due to the geometrical dependence of the flux, the perturbation in the
\Lya flux due to biased density fluctuations is a linear functional of
the underlying density fluctuation field, i.e., the resulting
contribution to $\td_{\alpha} (\bk)$ is related to the Fourier
transform of the total density perturbation $\td_{\rm tot} (\bk)$ by
multiplication by an effective window function $\tilde{W}(k)$
\citep{BL05b}. The total density perturbation is defined as the
mass-weighted mean of the dark matter and baryon perturbations; we
also define $r_{[\delta:\delta_b]} \equiv \td_{\rm tot}/\td_b$.  Under
the conditions considered in this paper, the three observable power
spectra (eq.~\ref{powTb}) can be used to study separately the two
sources of \Lya fluctuations. In particular, the quantity \beq P_{\rm
flux-\delta}(k) \equiv P_{\mu^2}- \frac{2 \beta}
{r_{[\dot{\delta}:\delta]}} P_{\mu^4} = 2 r_{[\dot{\delta}:\delta]}
\beta_\alpha r_{[\delta:\delta_b]}\, \tilde{W}(k)\, P_{\delta_b}(k)
\label{eq:Pflux} \eeq 
is proportional to the biased-density contribution to $P_{\alpha}$,
while $P_{\rm un-\delta}(k)$ equals the Poisson contribution to
$P_{\alpha}$ except for a factor of $\beta^2_\alpha$. 

% In addition to these quantities, $P_{\mu^0}$ yields the (theoretically
% known) baryonic density power spectrum at each redshift, and its
% normalization (measured in mK) thus yields the mean brightness
% temperature $\bar{T}_b$.

\section{Observable signature}

The coupling of the spin temperature to the kinetic temperature
through the WF effect requires a relatively low cosmic \Lya
background, and is expected to occur well before the end of cosmic
reionization \citep{madau}. We illustrate our results assuming that
the \Lya coupling transition achieves $x_\alpha=1$ at $z=20$; this
requirement determines the star formation efficiency. We assume that
galaxies form within all dark matter halos above some minimum mass
(or, equivalently, a minimum circular velocity $V_c = \sqrt{GM/R}$ in
terms of the virial radius $R$). We consider a wide range of possible
values: galaxies that form through molecular hydrogen cooling ($V_c =
4.5$ km/s), atomic cooling ($V_c = 16.5$ km/s), or a minimum mass ten
times larger ($V_c = 35.5$ km/s) due to feedback effects in low-mass
halos. We consider a single H~II region size (calculated as having the
flux-weighted mean volume); this size is affected by the fraction
$f_{\rm esc}$ of stellar ionizing photons that escape into the IGM,
and by the stellar population. We consider two extremes, a stellar
initial mass function as observed locally (Pop II) or that expected
for the very first stars (Pop III; 100 solar mass, zero-metallicity
stars).

\begin{figure}
\includegraphics[width=84mm]{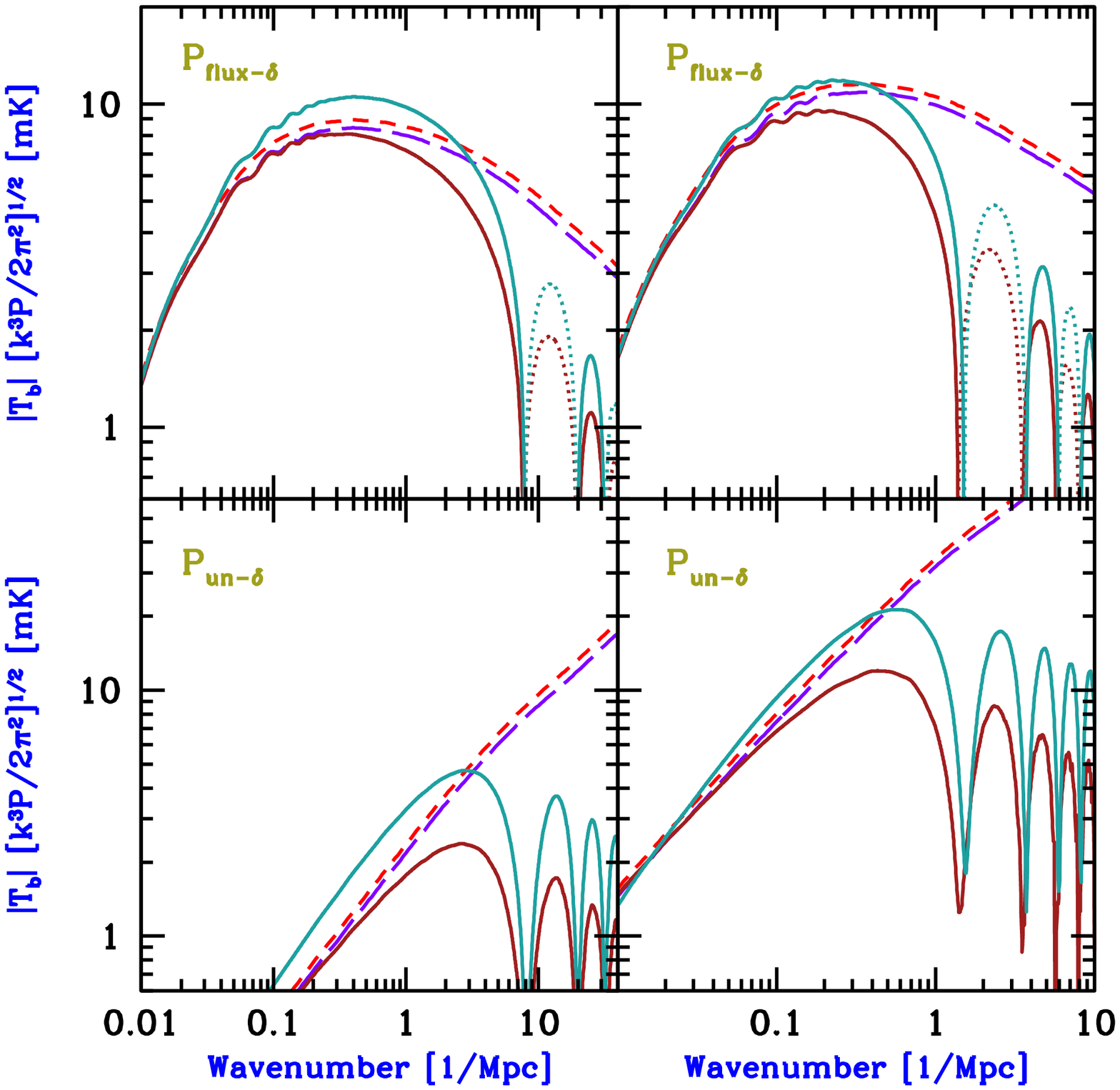}
\caption{21-cm power spectrum $P$ (shown in terms of the brightness
  temperature fluctuation) as a function of the comoving wavenumber
  $k$. We consider $P_{\rm flux-\delta}$ (top panels) and $P_{\rm
  un-\delta}$ (bottom panels). We compare the previous result
  \citep{BL05a,Jonathan06a,Hirata} (short-dashed curves) and the
  result corrected to use the precise density and temperature power
  spectra \citep{NB} (long-dashed curves) to the full calculation with
  the H~II region cutoff (solid curves, with negative portions dotted
  and shown in absolute value; the higher curve of each pair also
  includes the redistribution of photons due to scattering). We
  consider galactic halos with a minimum circular velocity $V_c =
  16.5$ km/s (left panels) or 35.5 km/s (right panels),
  assuming Pop II stars with $f_{\rm esc}=0.3$.  }
\label{PS}
\end{figure}

In Figure~\ref{PS} we show in two representative cases the two
separately-measurable power spectrum terms that are due to \Lya flux
fluctuations arising from biased density fluctuations ($P_{\rm
flux-\delta}$) or from Poisson fluctuations ($P_{\rm
un-\delta}$). Each term ($P_{\rm flux-\delta}$ or
$P_{\rm un-\delta}$) starts small on large scales and rises with $k$,
forming a peak before dropping (and then oscillating) on the scale of
the H~II region. Scattering modifies $P_{\rm un-\delta}$ more
substantially than $P_{\rm flux-\delta}$, since the former is more
strongly dominated by fluctuations on small scales. Note that our
assumption of a single H~II region size is reasonable since galaxies
at these redshifts are rare and almost all lie in halos close to the
minimum halo mass. Also, sufficiently early on in reionization, galaxy
clustering plays a relatively minor role in determining bubble
sizes. In particular, in the cases shown in Figure~\ref{PS}, the
cosmic ionized fraction is $\sim 10^{-3}$, and bubbles of sizes
between the minimum size (that of a single-source bubble) and twice
that size contain most of the ionized volume [2/3 for $V_c = 16.5$
km/s, and 1/2 for $V_c = 35.5$ km/s, using the \citet{FZH}
model]. While the small-scale ringing seen in Figure~\ref{PS} may be
smoothed out by a mild scatter in H~II region sizes, the overall shape
and in particular the peak of each power spectrum term are robustly
determined.

\begin{figure}%[th]
\includegraphics[width=84mm]{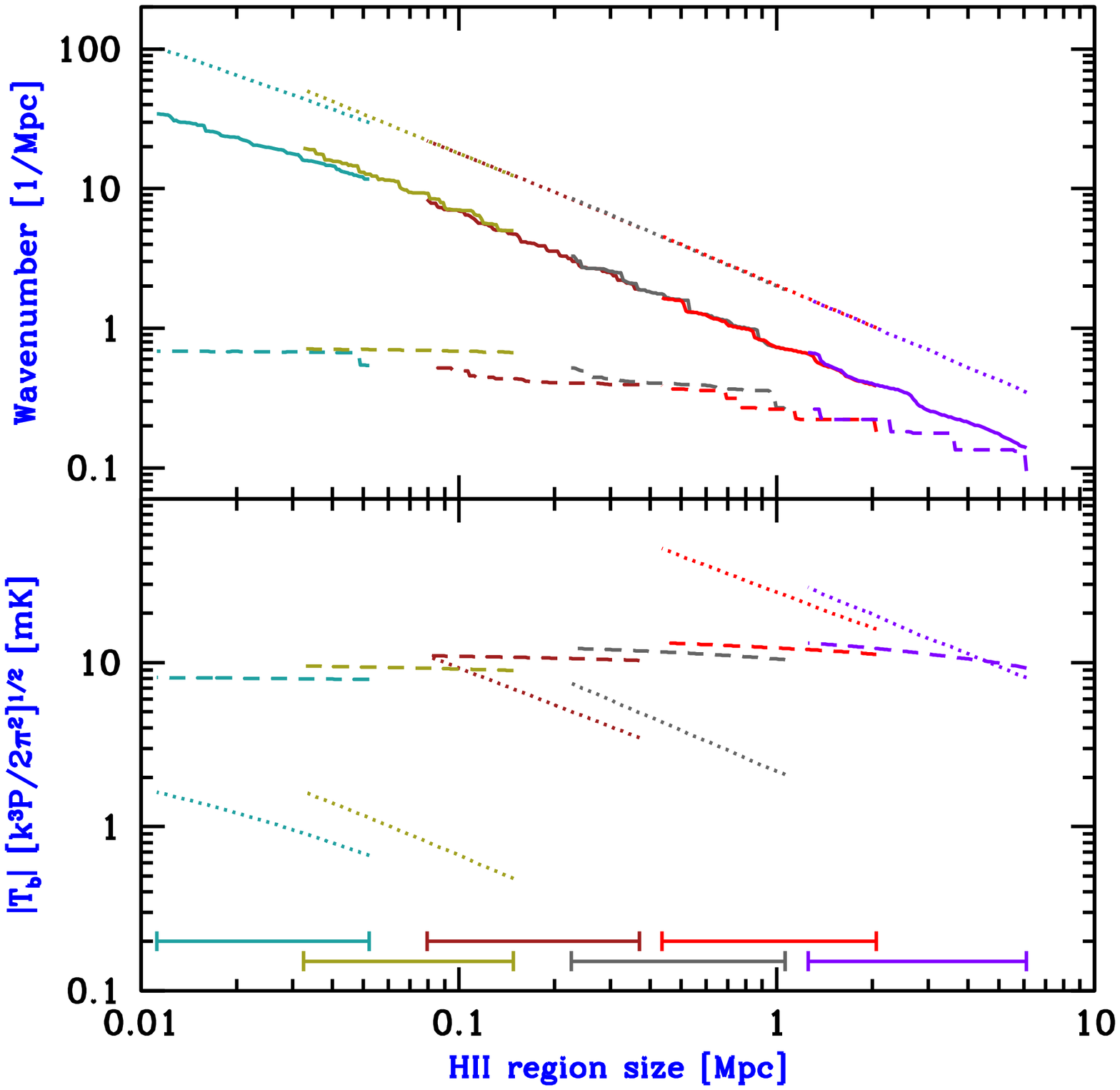}
\caption{Main features of 21-cm power spectra from \Lya flux
  fluctuations as a function of the comoving H~II region size. We show
  (top panel) $k_{\rm peak}^{\rm flux-\delta}$ (dashed curves),
  $k_{\rm peak}^{\rm un-\delta}$ (solid curves), and $k_{\rm 1\,
  mK}^{\rm flux-\delta}$ (dotted curves), as well as (bottom panel)
  $T_{\rm peak}^{\rm flux-\delta}$ (dashed curves) and $T_{\rm
  peak}^{\rm un-\delta}$ (dotted curves); see text for
  definitions. These curves are the result of a sweep of the parameter
  space, made in six segments each of which varies $f_{\rm esc}$ from
  1--$100\%$ thus covering a range of H~II region sizes shown by one
  of the horizontal bars (bottom panel): $V_c = 4.5$ km/s with Pop II
  stars or Pop III stars, $V_c = 16.5$ km/s with Pop II or III, and
  $V_c = 35.5$ km/s with Pop II or III stars, from left to right.}
\label{Mf}
\end{figure}

We summarize the main features of the two power spectra through the
peak positions ($k_{\rm peak}^{\rm flux-\delta}$ and $k_{\rm
peak}^{\rm un-\delta}$) and heights ($T_{\rm peak}^{\rm flux-\delta}$
and $T_{\rm peak}^{\rm un-\delta}$). Now, since the flux term would
have a peak even without the H~II region cutoff \citep{BL05b}, the
characteristics of its peak are fairly insensitive to the H~II region
size; thus, we also consider the position of its cutoff, specifically
the lowest $k$ value (above $k_{\rm peak}^{\rm flux-\delta}$) where
the power spectrum drops to 1 mK: $k_{\rm 1\,mK}^{\rm
flux-\delta}$. Both $k_{\rm peak}^{\rm un-\delta}$ and $k_{\rm
1\,mK}^{\rm flux-\delta}$ essentially measure the size $R_{\rm H II}$
of the H~II region (figure~\ref{Mf}). In particular, the product
$k_{\rm peak}^{\rm un-\delta} * R_{\rm H II} \approx 0.6$ and $k_{\rm
1 mK}^{\rm flux-\delta}* R_{\rm H II} \approx 1.6$, each to within a
factor of 1.5 over a range of three orders of magnitude of $R_{\rm H
II}$. On the other hand, the position and height of the peak of
$P_{\rm flux-\delta}$ are relatively insensitive to $R_{\rm H II}$ and
thus observing them would constitute a clear consistency check with
the theory. The Poisson peak height $T_{\rm peak}^{\rm un-\delta}$
measures the average number density of galaxies and thus depends
mainly on the minimum mass of galactic halos.

We note that if significant X-ray heating happened to occur
simultaneously with the \Lya coupling transition, then similar
fluctuations in the X-ray flux from galaxies would generate 21-cm
fluctuations that are even somewhat larger \citep{Jonathan07}; while
not included in previous studies, the X-ray fluctuations should
similarly be enhanced by scattering and should also show a cutoff at
the H~II bubble scale. While X-ray heating could possibly occur early,
\Lya heating is insignificant except after the \Lya
coupling fully saturates \citep{Miralda}.

The five quantitative characteristics we have focused on typically
occur at relatively large scales -- a fraction of a Mpc, or $\sim10$
arcseconds at $z=20$ -- and correspond to relatively large (but still
linear) fluctuations: 1--10 mK on a mean background of $\bar{T}_b \sim
-100$ mK. Thus, the predictions are theoretically robust, and require
observational capabilities only somewhat beyond those of the radio
arrays being currently constructed. Given current capabilities, 21-cm
cosmology is the most promising method for firmly detecting and
studying the properties of some of the earliest galaxies that ever
formed.

\section*{Acknowledgments}
We acknowledge support by Israel Science Foundation grant 629/05 and
U.S. - Israel Binational Science Foundation grant 2004386.

%\bsp

\label{lastpage}

\end{document}